\documentclass[thmsa,12pt]{article}
\usepackage{amsfonts}
\usepackage{amssymb}
\usepackage{amsmath}

\numberwithin{equation}{section}

\begin{document}


\title{Attractor horizons in six-dimensional \\ type IIB supergravity}

\author{Dumitru Astefanesei$\,^a$, Olivera Miskovic$\,^a$, and Rodrigo Olea$\,^b$\medskip \\
$^a$\emph{{\small Instituto de F\'\i sica, Pontificia Universidad Cat\'olica de Valpara\'\i so, }}\\
\emph{{\small Casilla 4059, Valpara\'{\i}so, Chile.}}\\
$^b$\emph{{\small Universidad Andres Bello, Departamento de Ciencias F\'\i sicas, }}\\
\emph{{\small Rep\'ublica 220, Santiago, Chile.\medskip }}\\
{\small dumitru.astefanesei@ucv.cl, olivera.miskovic@ucv.cl, }\\
{\small rodrigo.olea@unab.cl}}

\maketitle

\begin{abstract}

We consider near horizon geometries of extremal black holes in six-dimensional type
IIB supergravity. In particular, we use the entropy function formalism to compute the
charges and thermodynamic entropy of these solutions. We also comment on the role of
attractor mechanism in understanding the entropy of the Hopf T-dual solutions in type
IIA supergravity.

\end{abstract}

\newpage

\section{Introduction}

Warped anti-de Sitter spacetimes in three dimensions (WAdS$_3$) and their quotients play an important role in the understanding of non-supersymmetric extremal black holes entropy. In particular, WAdS$_3$ geometries are vacuum solutions of Topologically Massive Gravity, whose stability has been discussed in Ref.\cite{arXiv:0905.2612}. Thus, black hole solutions that are asymptotically WAdS$_3$ are also expected to exist. Indeed, these black holes are obtained as discrete quotients of WAdS$_3$ in Ref.\cite{arXiv:0807.3040} in the same way the BTZ black hole \cite{ hep-th/9204099, gr-qc/9302012} is a quotient space of AdS$_3$. Furthermore, WAdS$_3$ appears in the near horizon geometry of spinning extremal black holes and its role is essential in realization of the Kerr/CFT correspondence \cite{Guica:2008mu} (see, also, \cite{arXiv:1103.2355} and the references therein).

An important question is whether there is a way to embed WAdS$_3$ and its quotients in string/ supergravity (SUGRA) theories\footnote{WAdS$_3$ can be lifted to a full string theory solution in ten dimensions \cite{arXiv:0808.1912} (see, also,
\cite{Colgain:2010rg} for a related discussion).} without a gravitational Chern-Simons term in three dimensions. Once this can be done, one can explore the similarities with the better understood AdS$_3$ case and try to understand the properties of their dual conformal field theories (CFTs). For example, one can obtain important information about the dual CFT, such as the central charge, by studying the corresponding black holes in the bulk gravity.

Interestingly enough, this embedding can be obtained via the near horizon geometries of extremal black holes  \cite{Duff:1998cr} (see, also, Ref.\cite{Bena:2012wc}) -- a discussion of some properties of the dual CFT of these black holes was presented in a nice paper of Anninos \cite{Anninos:2008qb}.

In this letter, we generalize some of the results of Refs.\cite{Duff:1998cr, Anninos:2008qb} by turning on all moduli. In particular, we explicitly obtain the near horizon geometries of extremal solutions in six-dimensional type IIB SUGRA. We start with a generic near horizon geometry  WAdS$_3$$\times S^3$ and show that the physical solutions exist only for a particular fibration of WAdS$_3$ that, in fact, corresponds to AdS$_3$. This result is similar to the one obtained in Refs.\cite{Duff:1998cr, Anninos:2008qb}, where the moduli are turned off. Since these solutions are related by Hopf T-duality \cite{Duff:1998cr} to solutions of type IIA SUGRA with an WAdS$_3$ in the near horizon geometry \cite{noi}, we expect that their study would shed some light on the properties of their ``cousins'' in type IIA SUGRA.

When the axions are turned on, the entropy function has flat directions and one may wonder if the arguments of Ref.\cite{Anninos:2008qb} are still valid in this case. However, it was proven in Ref.\cite{ef3} that, even if the
near horizon background is not uniquely determined by the extremization equations, the entropy is still independent of the asymptotic data. Therefore, due to the attractor mechanism, the results of Ref.\cite{Anninos:2008qb} can be generalized in this case. That is, since the entropies of warped and unwarped solutions match, states with vanishing right-moving temperature in the dual CFT of type IIB
solution can still be mapped to thermal states with vanishing right-moving temperature in the dual CFT
of type IIA solution.

\section{Setup}

Instead of looking at the full dimensional reduction of the type IIB action on a six-dimensional torus $T^6$ \cite{Cremmer:1997ct}, we focus on a
subset of fields \cite{Duff:1998cr} for which one can obtain general enough solutions. The action constructed as a consistent truncation of IIB string theory to six dimensions has the form \cite{Duff:1998cr}
\begin{eqnarray}
I_{6B} &=&\int d^6x\,\sqrt{-G}\,\mathcal{L}_{6B}  \notag \\
&=&\frac{1}{16\pi G_{6}}\int d^6x\,\sqrt{-G}\,\left[ R-\frac{1}{2}\,\left(\partial \phi_1\right)^2-\frac{1}{2}
\,\left( \partial \phi_2\right) ^2-\frac{1}{2}\,e^{2\phi_1}\left( \partial \chi_1\right)^2\right.   \notag \\
&&\qquad \left. -\frac{1}{2}\,e^{2\phi_2}\left(\partial \chi_2\right)^2-\frac{1}{12}
\,e^{-\phi_1-\phi_2}F^2-\frac{1}{12}\,e^{\phi_1-\phi_2}K^2+\frac{1}{6}\,\chi _{2}\,F\left. ^{\ast }H\right.\right] \,,
\label{IIB action}
\end{eqnarray}
where the six-dimensional manifold is endowed by the metric $G_{\mu \nu }(x)$ and the corresponding scalar curvature is $R$. The theory contains four
scalar fields --two dilatons $\phi _1$, $\phi _2$ and two axions $\chi_1$, $\chi_2$,-- the gauge field 2-form $A$ in the NS sector of the
theory with the associated field strength 3-form $F=dA$, and also the RR gauge field 2-form $B$ whose field strength 3-form is $K=dB+\chi _1\,dA$. We have also introduced a 3-form $H=dB$ and defined a dual 3-form whose components are
\begin{equation}
^{*}H^{\mu \nu \lambda }=\frac{1}{3!}\frac{1}{\sqrt{-G}}\,\epsilon ^{\mu \nu
\lambda \alpha \beta \gamma }H_{\alpha \beta \gamma }\,,
\end{equation}
where $\epsilon^{\mu \nu \lambda \alpha \beta \gamma }$ is the constant Levi-Civit\`{a} tensor density. In our notation the indices are contracted as $FH=F_{\mu\nu\lambda }H^{\mu\nu\lambda }$ and, similarly, $F\left.^\ast H\right. =F_{\mu\nu\lambda }\left.^* H^{\mu\nu\lambda}\right.$.

The equations of motion derived from the action $I_{6B}$ are presented in Appendix A.
Since this action is both diffeomorphism and gauge invariant, one can apply the entropy function formalism \cite{ef3,ef1,ef2} to obtain the near horizon data of extremal black hole solutions of the theory. The near horizon geometry can be described by a line element that has the form of a product space WAdS$_3\times S^3$, and so we consider the following ansatz:
\begin{equation}
ds^2=G_{\mu \nu }\,dx^\mu dx^\nu =v_1\left( -\rho ^2dt^2
+\frac{d\rho ^2}{\rho ^2}\right) +u\left( d\varphi +e\rho \,dt\right)^2+v_2\,d\Omega _3 ^2\,,  \label{metric}
\end{equation}
where the local coordinates of the metric of WAdS$_3$  black hole near the horizon are $t\in \mathbb{R}$, $\rho =r-r_h\in \lbrack
0,\infty )$ and $\varphi \in \lbrack 0,2\pi )$, so that the horizon $r=r_{h}$ corresponds to $\rho =0$. The parameters $v_i $, $u>0$, and $e$ are constant. The Ricci tensor $^3R_{\ \nu}^\mu $ and scalar curvature $^3 R$ of this three-dimensional submanifold
with the coordinates $(t,\rho ,\varphi )$ have the form
\begin{equation}
\text{WAdS}_3:\quad^3 R_{\ \nu }^\mu =\left(\begin{array}{ccc}-\frac{2v_1-ue^2}{2v_1^2} & 0 & 0 \\
0 & -\frac{2v_1-ue^2}{2v_1^2} & 0 \\
\frac{e\rho \,\left( v_1-ue^2\right) }{2v_1^2} & 0 & -\frac{ue^2}{2v_1^2}\end{array}\right)
\,,\quad^3 R=\frac{ue^2-4v_1}{2v_1^2}\,,
\end{equation}
where the parameters $v_1$, $u$, and $e$ are arbitrary. The particular case for which $v_1=ue^2$ corresponds to a spacetime of
constant negative curvature with the AdS$_3$ radius $\ell =2\sqrt{v_1}$.

For a 3-sphere $S^3 $ of size $v_2$, we choose spherical coordinates $\psi ,\theta \in \lbrack 0,\pi )$ and $\alpha \in \lbrack 0,2\pi )$, such
that
\begin{eqnarray}
d\Omega_3 ^2 &=&d\psi ^2+\sin ^2\psi \,\left( d\theta ^2+\sin^2\theta \,d\alpha ^2\right) \,, \\
\varepsilon (S^3 ) &=&\sin ^2\psi \sin \theta \,d\psi \wedge d\theta\wedge d\alpha \,, \\
\text{Vol}(S^3 ) &=&\int \varepsilon (S^3 )=2\pi ^2\,.
\end{eqnarray}

The metric $G_{\mu \nu }$ in the local coordinates $x^\mu =\left( t,\rho,\varphi ,\psi ,\theta ,\alpha \right) $ then reads
\begin{equation}
G_{\mu \nu }=\text{diag}\left( \left[\begin{array}{ccc}-\left( v_1-ue^2\right) \rho ^2 & 0 & eu\rho \\
0 & \frac{v_1}{\rho ^2} & 0 \\
eu\rho & 0 & u\end{array}
\right] ,\left[
\begin{array}{ccc}
v_2 & 0 & 0 \\
0 & v_2\sin ^2\psi & 0 \\
0 & 0 & v_2\sin ^2\psi \sin ^2\theta
\end{array}
\right] \right) \,,  \label{G}
\end{equation}
so that the invariant volume element is constructed with $\sqrt{-G}=u^{1/2}v_1\,v_2^{3/2}\sin ^2\psi \sin \theta $, and the inverse
metric $G^{\mu \nu }$ is
\begin{equation}
G^{\mu \nu }=\text{diag}\left( \left[ \begin{array}{ccc}
-\frac{1}{v_1\rho ^2} & 0 & \frac{e}{v_1\rho } \\
0 & \frac{\rho ^2}{v_1} & 0 \\
\frac{e}{v_1\rho } & 0 & \frac{v_1-ue^2}{uv_1}\end{array} \right] ,\ \left[
\begin{array}{ccc} \frac{1}{v_2} & 0 & 0 \\
0 & \frac{1}{v_2\sin ^2\psi } & 0 \\
0 & 0 & \frac{1}{v_2\sin ^2\psi \sin ^2\theta }
\end{array} \right] \right) \,.
\end{equation}

In order to adopt an ansatz for the gauge fields, let us note that the 3-form $K=dB+\chi _1\,dA$ has been defined so that the action explicitly exhibits original symmetries of the eleven-dimensional supergravity \cite{Cremmer:1997ct}, but it requires an additional constraint that imposes the Bianchi identity on $K$.  We shall, however, work with the original tensor field strengths $F=dA$  and $H=dB$ that automatically satisfy the Bianchi identities and, in terms of them, $K=H+\chi _1\,F$.

An ansatz for the field strengths $F$ and $H$ that possesses the same symmetries as the metric $G_{\mu \nu }$ (that is, $\pounds _{\xi _{(i)}}F=0$ and
$\pounds _{\xi_{(i)}}H=0$), can be written as (for details, see Appendix B)
\begin{eqnarray}
F &=&e_1\,dt\wedge d\rho \wedge d\varphi +p_1\,\varepsilon (S^3)\,,
\label{F} \\
H &=&e_2\,dt\wedge d\rho \wedge d\varphi +p_2\,\varepsilon (S^3)\,,
\label{H}
\end{eqnarray}
where $e_i$ and $p_i$ are electric and magnetic charges, respectively.

The ansatz (\ref{metric}), (\ref{F}), and (\ref{H}) is similar to the one used in Ref. \cite{Dabholkar:2006tb} (see, also, Refs.\cite{arXiv:1008.3852,arXiv:1104.4121} for similar considerations in AdS).

\section{Entropy function}
To get the near horizon data, we employ the entropy function formalism presented in Refs.
\cite{ef3,ef1,ef2} (see, also, Ref. \cite{arXiv:0708.1270}).

We want to calculate the entropy function, $\mathcal{E}$, defined by
\begin{equation}
\mathcal{E}\left( \vec{u},\vec{v},\vec{e},\vec{p}\right) =2\pi \left( \rule{0pt}{18pt}q_i e_i-f\left( \vec{u},\vec{v},\vec{e},\vec{p}\right) \right)
\,,  \label{E-function}
\end{equation}
where the electric charges are $\vec{q}=\partial f/\partial \vec{e}$, and
\begin{equation}
f=\int\limits_{\mathcal{H}} d^4 x\,\sqrt{-G}\,\mathcal{L}_{\text{6B}}
\label{f}
\end{equation}
is the action evaluated on the horizon $\mathcal{H}$ on the background (\ref{metric}), (\ref{F}) and (\ref{H}) for fixed time.
In $\mathcal{L}_{\text{6B}}$, all fields are evaluated at the horizon, as well. In particular, the values of the moduli (scalar fields) at the horizon correspond to a set of parameters $\phi_1(r_h)=u_1$, $\phi_2(r_h)=u_2$, $\chi_1(r_h)=u_3 $ and $\chi_2(r_h)=u_4$. Explicitly, we can write
\begin{equation}
f=\frac{1}{16\pi G_{6}}\,\int\limits_{\mathcal{H}}d\varphi d\psi d\theta d\alpha \,\sqrt{-G}\,\left[ R-\frac{1}{12}\,e^{-u_1-u_2}F^2
-\frac{1}{12}\,e^{u_1-u_2}\left( H+\chi _1\,F\right)^2+\chi_2\,F\wedge H\right] \,,
\end{equation}
which, further, taken on a  near horizon background, becomes
\begin{eqnarray}
f &=&\frac{\pi }{8G_{6}}\,u^{1/2}v_{1}\,v_{2}^{3/2}\left[ \frac{12}{v_{2}}-\frac{4}{v_{1}}+\frac{ue^{2}}{v_{1}^{2}}+e^{-u_{1}-u_{2}}
\left( \frac{e_{1}^{2}}{uv_{1}^{2}}-\frac{p_{1}^{2}}{v_{2}^{3}}\right) \right.   \notag \\
&&\left. +e^{u_{1}-u_{2}}\left( \frac{\left( e_{2}+u_{3}e_{1}\right) ^{2}}{uv_{1}^{2}}
-\frac{\left( p_{2}+u_{3}p_{1}\right) ^{2}}{v_{2}^{3}}\right) \right] +\frac{\pi }{4G_{6}}\,u_{4}\,\left( e_{1}p_{2}-e_{2}p_{1}\right) \,.
\label{f_explicit}
\end{eqnarray}

The extremization of the entropy function $\mathcal{E}\left( \vec{u},\vec{v},\vec{e},\vec{p}\right) $ determines the parameters
\begin{equation}
\vec{u}=\{u,u_1,u_2,u_3,u_4 \}\ ,\quad \vec{v}=\{v_1,v_2\}\ ,\quad \vec{e}=\{e,e_1,e_2\}\ ,
\end{equation}
in terms of the electric charge $\vec{q}$ and magnetic charge $\vec{p}$ of an extremal black hole,
\begin{equation}
\vec{q}=\{q,q_1,q_2\}\ ,\quad \vec{p}=\{p_1,p_2\}\ ,
\end{equation}
so that, when the attractor mechanism holds, a value of the entropy $S=\mathcal{E}_{\text{ext}}$ does not dependent on the asymptotic values of the scalar fields.

The function $\mathcal{E}$, where $f$ is given by Eq.(\ref{f_explicit}), has an extremum for fixed charges $\vec{q}$, $\vec{p}$\textbf{\ }when
\begin{equation}
\frac{\partial f}{\partial \vec{u}}=0\,,\quad \frac{\partial f}{\partial \vec{v}}=0\,,\quad \frac{\partial f}{\partial \vec{e}}=\vec{q}\,.
\end{equation}

We solve the attractor equations, which are explicitly given in Appendix A, as follows.

Taking the sum and difference of Eqs.(\ref{U1}) and (\ref{U2}), we obtain $e_{1}$ and $e_{2}$,
\begin{equation}
\frac{e_{1}^{2}}{uv_{1}^{2}}=\frac{p_{1}^{2}}{v_{2}^{3}}\,,\qquad \frac{e_{2}^{2}}{uv_{1}^{2}}=\frac{p_{2}^{2}}{v_{2}^{3}}\,,  \label{selfduality}
\end{equation}
which means that $F$ and $H$ are either self-dual or anti self-dual field
strengths, $^*F=\mp F$ and $^*H=\mp H$, with the dual 3-forms having the components
\begin{equation}
^*F=-\frac{p_{1}u^{1/2}v_{1}}{v_{2}^{3/2}}\,dt\wedge d\rho \wedge d\varphi -\frac{e_{1}v_{2}^{3/2}}{u^{1/2}v_{1}}\,\varepsilon (S^3)\,,
\end{equation}
and similarly for $^*H$. Eq.(\ref{U3}) determines that $F$ and $H$ must be \textit{simultaneously} dual or anti self-dual. Then, Eq.(\ref{U4}) is identically satisfied.

The difference of Eqs.(\ref{U}) and (\ref{V1}) leads to a such $u$ that a near horizon geometry in the considered case is always AdS$_3$,
\begin{equation}
v_{1}=ue^{2}\,.  \label{u}
\end{equation}
The sum of Eqs.(\ref{V1}) and (\ref{V2}) relates the sizes of two product spaces AdS$_2$ and $S^3$ as
\begin{equation}
v_{2}=4v_{1}\, , \label{v2}
\end{equation}
as expected in higher dimensions; only in four dimensions and  for a vanishing moduli potential 
the radii of AdS$_2$ and the sphere are equal.

With the last two results, any of Eqs.(\ref{U}-\ref{V2}) gives the size of the AdS$_2$ space,
\begin{equation}
\left( 8v_{1}\right) ^{2}=e^{-u_{1}-u_{2}}\,p_{1}^{2}+e^{u_{1}-u_{2}}\left(p_{2}+u_{3}\,p_{1}\right) ^{2}\,.  \label{v1^2}
\end{equation}
Equations (\ref{E1}) and (\ref{E2}) then determine the dilatons,
\begin{eqnarray}
e^{-2u_{2}} &=&\left( \frac{4G_{6}}{\pi ^{2}}\frac{q_{1}-u_{3}q_{2}}{p_{2}+u_{3}p_{1}}-u_{4}\right)
\left( \frac{4G_{6}}{\pi ^{2}}\frac{q_{2}}{p_{1}}+u_{4}\right)   \label{u2} \\
e^{-2u_{1}} &=&\frac{p_{2}+u_{3}p_{1}}{p_{1}\left( \frac{4G_{6}}{\pi ^{2}}\,q_{2}+u_{4}p_{1}\right) }\left[ \frac{4G_{6}}{\pi ^{2}}\,
\left(q_{1}-u_{3}q_{2}\right) -u_{4}\left( p_{2}+u_{3}p_{1}\right) \right] ,
\end{eqnarray}
where $p_{1}\neq 0$, $u_{3}\neq -\frac{p_{2}}{p_{1}}$ and $u_{4}\neq -\frac{4G_{6}}{\pi ^{2}}\,\frac{q_{2}}{p_{1}}$. The values of axions on the
horizon remain arbitrary, which means that the entropy function has flat directions. However, as expected, the size of AdS space (\ref{v1^2}) does not depend on them,
\begin{equation}
v_{1}=\sqrt{\pm \frac{G_{6}}{16\pi ^{2}}\,\left(p_{1}q_{1}+p_{2}q_{2}\right) }\,.  \label{v1}
\end{equation}
Finally, the last unknown parameter is found from Eq.(\ref{E})
\begin{equation}
e=\sqrt{\pm \frac{p_{1}q_{1}+p_{2}q_{2}}{8q}}\,.  \label{e}
\end{equation}
The signs must be chosen so that $v_{1}^{2}$ and $e^{2}$ are positive. Thus, depending on the signs of the electric $q_{i}$ and magnetic $p_{i}$ charges,
only one (dual or anti self-dual) solution is admissible.

To summarize, the entropy function $\mathcal{E}$ has an extremum for the non-trivial values of the parameters $\vec{u}$, $\vec{v}$ and $\vec{e}$
expressed in terms of $\vec{p}$ and $\vec{q}$ by equations (\ref{u}), (\ref{v2}), (\ref{u2} -- \ref{e}) and
\begin{equation}
e_{1}=\pm \dfrac{p_{1}}{8e}\,,\qquad e_{2}=\pm \dfrac{p_{2}}{8e}\,.
\end{equation}

It is important to emphasize that, since $v_{1}=ue^{2}$, the \emph{warped geometry is not allowed} for near horizon solutions in type IIB supergravity when there exists spherical symmetry in the other three angular directions. In the near horizon limit, the $3$-forms $F$ and $H$ become (anti) self-dual. Note also that, as expected, the entropy does not depend on the flat directions and it is a function of the physical charges only,
\begin{equation}
\mathcal{E}_{\text{ext}}=4\pi qe=\sqrt{\pm 2\pi ^{2}\,(p_{1}q_{1}+p_{2}q_{2})q}\,.
\end{equation}

When all scalar fields (dilatons and axions) vanish, the moduli equations of the full action (\ref{IIB action}) become (see Appendix \ref{E.o.m.})
\begin{eqnarray}
F^2 &=&0\,,\qquad FH=0\,, \\
H^2 &=&0\,,\qquad F\left. ^*H\right. =0\,,
\end{eqnarray}
whose only solutions are dual and self-dual gauge field configurations. This is consistent with the results of Ref.\cite{Duff:1998cr}.

\section{Discussion}
In this letter, we have explicitly constructed near horizon attractor geometries in
type IIB SUGRA theory and generalized some results of Ref.\cite{Duff:1998cr} in the presence of
the axions. These configurations are important because they can be related by Hopf
T-dualities \cite{Duff:1998cr}\footnote{This method was also applied to Taub-Nut spacetimes
in Ref.\cite{Astefanesei:2005yj}.} to black hole solutions in type IIA SUGRA that have an WAdS$_3$
in the near horizon geometry, and also could be relevant to understanding the Kerr/CFT
correspondence.

Since we have used the entropy function formalism, we were able to exactly obtain
the near horizon data and macroscopic entropy. As explained in Section $3$, with
the axions turned on, the entropy function has flat directions and so the near horizon
data do not completely decouple from the asymptotic data of the scalar fields. At first sight, this
result could be problematic for the interpretations of the dual CFT of Hopf
T-dual type IIA black hole solutions \cite{Anninos:2008qb}. However, it is known
that, due to the long throat of AdS$_2$, the entropy is still independent of these
asymptotic data \cite{ef3}. A successful application of this method is an indication
that the attractor mechanism should also work for the Hopf T-dual solutions. More
importantly, the attractor mechanism guarantees
that even if the supersymmetry is broken by the Hopf T-duality transformation, one
can still compute the statistical entropy for which the microscopic theory
is weakly coupled \cite{Dabholkar:2006tb, Astefanesei:2006sy}.

A further consistent truncation of action (\ref{IIB action}) can be obtained by
turning off the axions and a black hole solution in this theory was presented in
\cite{Anninos:2008qb}. The advantage of using the entropy function formalism is
that we can explicitly obtain the values of the dilatons at the horizon, which
are
$$
e^{\phi_1(r_h)}=\sqrt{\frac{p_1q_2}{q_1p_2}}\,,
\qquad e^{\phi_2(r_h)}=\frac{\pi^2}{4G_{6}}\sqrt{\frac{p_1 p_2}{q_1 q_2}}\,.
$$
In this case, there are no flat directions and the
near horizon data are completely fixed by the electric and magnetic charges due to the
attractor mechanism.

Extremal spinning black holes have an WAdS$_3$ geometry near the horizon and, in principle, one can
Kaluza-Klein reduce the theory to obtain AdS$_2$ and then apply the
AdS$_2$/CFT$_1$ correspondence. Another proposal to compute the entropy of (non-SUSY)
extremal spinning black holes is Kerr/CFT correspondence, though, despite
some attempts to embed it in string theory \cite{arXiv:1009.5039}, it is
not on the same footing with the AdS/CFT duality.

Using the attractor mechanism and the universality of the near horizon geometry,
it was shown in \cite{Astefanesei:2009sh} that the central charge of the dual CFT of a large class
of extremal spinning black holes (computed from the Kerr/CFT correspondence) does
not match the central charge obtained from the AdS$_2$/CFT$_1$ (though, they are
proportional). Hence, even if the final entropy results match, the computations should
in fact correspond to two different embeddings in string theory.

Somehow similarly, one can use the observations of Refs.\cite{Duff:1998cr, Anninos:2008qb} to see that
AdS$_3$ and WAdS$_3$ also correspond to two different embeddings in string theory related
by Hopf T-duality. Since the attractor mechanism works in type IIB SUGRA, we expect
the same kind of universality of the entropy for non-SUSY extremal black hole solutions
(with an WAdS$_3$ in the near horizon geometry \cite{hep-th/0504231}) in type IIA SUGRA \cite{noi}.

\section*{Acknowledgments}
This work was funded by FONDECYT Grants \#1090357, \#1110102, and \#120446. O.M. thanks DII-PUCV for support through the project \#123.711/2011.
The work of R.O. is financed in part by the UNAB grant DI-117-12/R.

\appendix{}

\section{Equations of motion for the type IIB action } \label{E.o.m.}

A more general action of the type (\ref{IIB action}), describing six-dimensional gravity coupled to scalar fields $\phi^i$ and Abelian
gauge field 2-forms $A^A$ with the field-strength 3-forms $F^A=dA^A$,
is
\begin{eqnarray}
I[G,\phi ,A] &=&\frac{1}{16\pi G_{6}}\int d^6x\,\sqrt{-G}\,\left[ \rule{0pt}{18pt}R-\frac{1}{2}\,g_{ij}(\phi )\,\partial _\mu \phi^i \partial ^\mu \phi
^j \right.  \notag \\
&&\qquad -\left. \rule{0pt}{18pt}f_{AB}(\phi )\,F_{\mu \nu \lambda}^A F^{B\,\mu \nu \lambda }-\tilde{f}_{AB}(\phi )\,
F_{\mu \nu \lambda}^A \left. ^*F^{B\,\mu \nu \lambda }\right. \right] \,,
\end{eqnarray}
where the Hodge-dual field strength is $^*F^{B\,\mu \nu \lambda }=\frac{1}{3!\,\sqrt{-G}}\,\epsilon ^{\mu \nu \lambda \alpha \beta \gamma}F_{\alpha \beta \gamma }^{B}$
and the metric $G_{\mu\nu}(x)$ lowers and rise the spacetime indices. The equations of motion that extremize the above action are
\begin{eqnarray}
R_{\mu \nu }-\frac{1}{2}\,g_{ij}\,\partial _\mu \phi ^i \partial _\nu \phi ^j &=&f_{AB}\,\left( 3F_{\mu \alpha \beta }^A F_\nu ^{B\,\alpha \beta }-
\frac{1}{2}\,G_{\mu \nu }F_{\alpha \beta \gamma }^A F^{B\,\alpha \beta \gamma }\right) , \\
\frac{1}{\sqrt{-G}}\,\partial _\mu \left( \sqrt{-G}\,g_{ij}\,\partial ^\mu \phi ^j \right) &=&
\partial_i f_{AB}\,F_{\mu \nu\lambda }^A F^{B\,\mu \nu \lambda }+\partial_i \tilde{f}_{AB}
\,F_{\mu \nu \lambda }^A \left. ^*F^{B\,\mu \nu \lambda }\right.\,,
\end{eqnarray}
and
\begin{equation}
\partial _{\lambda }\left[ \sqrt{-G}\,\left( f_{AB}\,F^{B\,\mu \nu \lambda }+
\tilde{f}_{AB}\left. ^*F^{B\,\mu \nu \lambda }\right. \right) \right]=0\,,
\end{equation}
where we denote $\partial _i \equiv \partial /\partial \phi ^i $ and the
scalar curvature in the Einstein equation has been eliminated using $R=\frac{1}{2}\,g_{ij}(\phi )\,\partial _\mu \phi ^i \partial ^\mu \phi ^j $.

In particular, the field content of the action (\ref{IIB action}) is given by $\phi^i =\left\{\phi_1,\phi_2,\chi_1,\chi_2\right\}$ and
$F^A =\left\{ F,H\right\} $, with the interaction determined by
\begin{eqnarray}
g_{ij} &=&\text{diag}\left( 1,1,e^{2\phi _{1}},e^{2\phi _{2}}\right) \,, \\
f_{AB} &=&\frac{e^{\phi _{1}-\phi _{2}}}{12}\,\left(\begin{array}{cc}1+e^{-2\phi _{1}} & \chi _{1} \\
\chi _{1} & 1
\end{array}
\right) \,,\quad \tilde{f}_{AB}=\frac{\chi_2}{12}\left(\begin{array}{cc}
0 & -1 \\
1 & 0
\end{array}\right) \,.
\end{eqnarray}

The equations of motion in this case are as follows. The Einstein equations read
\begin{eqnarray}
R_{\mu \nu } &=&\frac{1}{2}\,\left( \partial _\mu \phi_1\partial _{\nu}\phi_1+\partial _\mu \phi_2\partial _\nu \phi_2+e^{2\phi
_1}\partial _\mu \chi_1\partial _\nu \chi_1+e^{2\phi_2}\partial _\mu \chi_2\partial _\nu \chi_2\right)   \notag \\
&&+\frac{1}{12}\,e^{-\phi_1-\phi_2}\left( 3F_{\mu \alpha \beta }F_{\nu}^{\ \alpha \beta }-\frac{1}{2}\,G_{\mu \nu }F^2\right)   \notag \\
&&+\frac{1}{12}\,e^{\phi_1-\phi_2}\left( 3K_{\mu \alpha \beta }K_{\nu}^{\ \alpha \beta }-\frac{1}{2}G_{\mu \nu }K^2\right) \,.
\end{eqnarray}
The field equations for dilations are
\begin{eqnarray}
\frac{1}{\sqrt{-G}}\,\partial _\mu \left( \sqrt{-G}\,G^{\mu \nu }\partial_\nu \phi_1\right) -e^{2\phi_1}\left( \partial \chi_1\right) ^2
&=&-\frac{1}{12}\,\left( e^{-\phi_1-\phi_2}F^2-e^{\phi_1-\phi_2}K^2\right) \,, \\
\frac{1}{\sqrt{-G}}\,\partial _\mu \left( \sqrt{-G}\,G^{\mu \nu }\partial_\nu \phi_2\right) -e^{2\phi_2}\left( \partial \chi_2\right) ^2
&=&-\frac{1}{12}\,\left( e^{-\phi_1-\phi_2}F^2+e^{-\phi_1-\phi_2}K^2\right) \,,
\end{eqnarray}
whereas for the axions they have the form
\begin{eqnarray}
\frac{1}{\sqrt{-G}}\,\partial _\mu \left( \sqrt{-G}\,G^{\mu \nu }e^{2\phi_1}\partial _\nu \chi_1\right)  &=&\frac{1}{6}\,e^{\phi_1-\phi_2}FK\,, \\
\frac{1}{\sqrt{-G}}\,\partial _\mu \left( \sqrt{-G}\,G^{\mu \nu }e^{2\phi_2}\partial _\nu \chi_2\right)  &=&\frac{1}{6}\,\,^*FH \,.
\end{eqnarray}
Finally, the equations of motion for the gauge fields $A_{\mu \nu }$ and $B_{\mu \nu }$ can be written as
\begin{eqnarray}
\frac{1}{\sqrt{-G}}\,\partial _\mu \left[ \sqrt{-G}\left( e^{-\phi_1-\phi _2}F^{\mu \nu \lambda }+\chi _1e^{\phi_1-\phi_2}K^{\mu
\nu \lambda }\right) \right]  &=&\partial _\mu \chi _2\left. ^*H^{\mu \nu \lambda }\right. , \\
\frac{1}{\sqrt{-G}}\,\partial _\mu \left( \sqrt{-G}\,e^{\phi_1-\phi_2}K^{\mu \nu \lambda }\right)
&=&-\partial_\mu\chi _2\,\left. ^*F^{\mu \nu \lambda }\right. .
\end{eqnarray}

In the near horizon limit when the background becomes (\ref{metric}), the attractor equations are
obtained by the extremization of the entropy function given by Eqs.(\ref{E-function}) and (\ref{f_explicit}). Varying it in the moduli gives
\begin{eqnarray}
\delta u_{1} &:&\quad 0=-e^{-u_{1}-u_{2}}\left( \frac{e_{1}^{2}}{uv_{1}^{2}}-\frac{p_{1}^{2}}{v_{2}^{3}}\right) +e^{u_{1}-u_{2}}\left(
\frac{e_{2}^{2}+u_{3}^{2}\,e_{1}^{2}}{uv_{1}^{2}}-\frac{p_{2}^{2}+u_{3}^{2}\,p_{1}^{2}}{v_{2}^{3}}\right) \,,  \label{U1} \\
\delta u_{2} &:&\quad 0=e^{-u_{1}-u_{2}}\left( \frac{e_{1}^{2}}{uv_{1}^{2}}-\frac{p_{1}^{2}}{v_{2}^{3}}\right) +e^{u_{1}-u_{2}}
\left( \frac{e_{2}^{2}+u_{3}^{2}\,e_{1}^{2}}{uv_{1}^{2}}-\frac{p_{2}^{2}+u_{3}^{2}\,p_{1}^{2}}{v_{2}^{3}}\right) \,,  \label{U2} \\
\delta u_{3} &:&\quad 0=u_{3}\,\left( \frac{e_{1}^{2}}{uv_{1}^{2}}-\frac{p_{1}^{2}}{v_{2}^{3}}\right) +\left( \frac{e_{1}e_{2}}{uv_{1}^{2}}
-\frac{p_{1}p_{2}}{v_{2}^{3}}\right) \,,  \label{U3} \\
\delta u_{4} &:&\quad 0=e_{1}p_{2}-e_{2}p_{1}\,,  \label{U4}
\end{eqnarray}
and varying it in the metric parameters $\delta u$, $\delta v_1$ and $\delta v_2$ gives, respectively,
\begin{eqnarray}
\frac{12}{v_{2}}-\frac{4}{v_{1}}+\frac{3ue^{2}}{v_{1}^{2}}
\hspace{-2mm}&=&\hspace{-2mm} e^{-u_{1}-u_{2}}\left( \frac{e_{1}^{2}}{uv_{1}^{2}}+\frac{p_{1}^{2}}{v_{2}^{3}}\right) +e^{u_{1}-u_{2}}
\left[ \frac{\left( e_{2}+u_{3}\,e_{1}\right) ^{2}}{uv_{1}^{2}}+\frac{\left( p_{2}+u_{3}\,p_{1}\right) ^{2}}{v_{2}^{3}}\right]\hspace{-1mm},   \label{U} \\
\frac{12}{v_{2}}-\frac{ue^{2}}{v_{1}^{2}}
\hspace{-2mm}&=& \hspace{-2mm} e^{-u_{1}-u_{2}}\left( \frac{e_{1}^{2}}{uv_{1}^{2}}+\frac{p_{1}^{2}}{v_{2}^{3}}\right)
+e^{u_{1}-u_{2}}\left[ \frac{\left( e_{2}+u_{3}\,e_{1}\right) ^{2}}{uv_{1}^{2}}+\frac{\left( p_{2}+u_{3}\,p_{1}\right) ^{2}}{v_{2}^{3}}\right]\hspace{-1mm},  \label{V1} \\
\frac{4}{v_{2}}-\frac{4}{v_{1}}+\frac{ue^{2}}{v_{1}^{2}}
\hspace{-2mm}&=&\hspace{-2mm}-e^{-u_{1}-u_{2}}\left( \frac{e_{1}^{2}}{uv_{1}^{2}}+\frac{p_{1}^{2}}{v_{2}^{3}}\right) \hspace{-1mm}-\hspace{-1mm}e^{u_{1}-u_{2}}
\left[ \frac{\left(e_{2}+u_{3}\,e_{1}\right) ^{2}}{uv_{1}^{2}}+\frac{\left(p_{2}+u_{3}\,p_{1}\right) ^{2}}{v_{2}^{3}}\right]\hspace{-1mm}.  \label{V2}
\end{eqnarray}
Finally, the extremization in $\vec{e}$ leads to
\begin{eqnarray}
\delta e_{1} &:&\quad q_{1}=\frac{\pi }{4G_{6}}\,\left[ e^{-u_{1}-u_{2}}\,\frac{e_{1}\,v_{2}^{3/2}}{v_{1}\,u^{1/2}}
+e^{u_{1}-u_{2}}\frac{\left( e_{2}+u_{3}e_{1}\right) u_{3}v_{2}^{3/2}}{v_{1}\,u^{1/2}}+u_{4}\,p_{2}\right] \,,  \label{E1} \\
\delta e_{2} &:&\quad q_{2}=\frac{\pi ^{2}}{4G_{6}}\,\left[ e^{u_{1}-u_{2}}\frac{v_{2}^{3/2}
\left(e_{2}+u_{3}e_{1}\right) }{v_{1}\,u^{1/2}}\,-u_{4}\,p_{1}\right] \,,
\label{E2} \\
\delta e &:&\quad q=\frac{\pi ^{2}}{4G_{6}
}\,
\frac{eu^{3/2}v_{2}^{3/2}}{v_{1}}\,.  \label{E}
\end{eqnarray}

\section{Near horizon isometries and invariant $p$-forms \label{Isometries}}

We want to find the ans\"{a}tze of $p$-forms
\begin{equation}
F_{(p)}=\frac{1}{p!}\,F_{\mu _1\cdots \mu _{p}}\,dx^{\mu _1}\wedge\cdots \wedge dx^{\mu _{p}}\,,\qquad p=2,3,4,
\end{equation}
that possess the same symmetries as a 6-dimensional spacetime that is a direct product of WAdS$_3$ and $S^3 $, as given by the metric $G_{\mu \nu
}$, Eq.(\ref{metric}). Since spacetime is a product space, the $p$-form with $p\leq 3$ is a sum of two $p$-forms, each one defined on a respective
three-dimensional subspace,
\begin{equation}
F_{(p)}=\left. F_{(p)}\right\vert _{\text{WAdS}_3 }+\left. F_{(p)}\right\vert _{S^3}\,.
\end{equation}

For this purpose, we find the isometries of $G_{\mu \nu }$. In the coordinates $x^\mu =(t,\rho ,\varphi ,\psi ,\theta ,\alpha )$ and for
$v_1-ue^2\neq 0$, the Killing equation
\begin{equation}
\pounds _{\xi }G_{\mu \nu }=\partial _\mu \xi ^{\lambda }G_{\lambda \nu}+\partial _\nu \xi ^{\lambda }G_{\mu \lambda }
+\xi ^{\lambda }\partial_{\lambda }G_{\mu \nu }=0\,,
\end{equation}
has $10$ linearly independent solutions $\xi ^{(i)}=\xi ^{(i)\mu }\partial_\mu $, that correspond\ to near horizon isometries
\begin{eqnarray*}
\xi ^{(1)} &=&\partial _{t}\,, \\
\xi ^{(2)} &=&t\partial _{t}-\rho \partial _{\rho }\,, \\
\xi ^{(3)} &=&-\left( t^2+\frac{1}{\rho ^2}\right) \partial _{t}+2t\rho\,\partial _{\rho }+\frac{2e}{\rho }\,\partial _{\varphi }\,, \\
\xi ^{(4)} &=&\partial _{\varphi }\,, \\
\xi ^{(5)} &=&-\cos \alpha \,\partial _{\theta }+\cot \theta \sin \alpha\,\partial _{\alpha }\,, \\
\xi ^{(6)} &=&\sin \alpha \,\partial _{\theta }+\cot \theta \cos \alpha
\,\partial _{\alpha }\,, \\
\xi ^{(7)} &=&\partial _{\alpha }\,, \\
\xi ^{(8)} &=&-\cos \theta \,\partial _{_{\psi }}+\cot \psi \sin \theta\,\partial _{\theta }\,, \\
\xi ^{(9)} &=&-\cos \alpha \sin \theta \,\partial _{_{\psi }}+\cot \psi
\left( -\cos \alpha \cos \theta \,\partial _{\theta }+\frac{\sin \alpha }{\sin \theta }\,\partial _{\alpha }\right) \,, \\
\xi ^{(10)} &=&\sin \alpha \sin \theta \,\partial _{_{\psi }}+\cot \psi
\left( \sin \alpha \cos \theta \,\partial _{\theta }+\frac{\cos \alpha }{\sin \theta }\,\partial _{\alpha }\right) \,.
\end{eqnarray*}

The isometry algebra $SL(2,\mathbb{R})\times U(1)\times SO(4)$ is generated
by the operators
\begin{equation*}
\begin{array}{lllll}
SL(2,\mathbb{R}): & \mathbf{L}_1=\frac{1}{2}\,\left( \xi ^{(1)}-\xi^{(3)}\right) \,, & SO(4): & \mathbf{J}_{12}=\xi ^{(7)}\,,
\qquad & \mathbf{J}_{23}=-\xi ^{(5)}\,, \\
& \mathbf{L}_2=\frac{1}{2}\,\left( \xi ^{(1)}+\xi ^{(3)}\right) \,,\qquad
&  & \mathbf{J}_{13}=-\xi ^{(6)}\,, & \mathbf{J}_{24}=-\xi ^{(9)}\,, \\
& \mathbf{L}_3 =\xi ^{(2)}\,,\medskip &  & \mathbf{J}_{14}=-\xi ^{(10)}\,,
& \mathbf{J}_{34}=\xi ^{(8)}\,, \\
U(1): & \mathbf{G}=\xi ^{(4)}\,, &  &  &
\end{array}
\end{equation*}
that satisfy the Lie brackets $\left[ \mathbf{L}_{k},\mathbf{L}_{l}\right] =\epsilon _{klm}\,\mathbf{L}^{m},$ and $\left[ \mathbf{J}_{ab},\mathbf{J}_{cd}\right] =\delta_{ad}\,\mathbf{J}_{bc}-\delta _{bd}\,\mathbf{J}_{ac}-\delta _{ac}\,\mathbf{J}_{bd}+\delta _{bc}\,\mathbf{J}_{ad}$.

Having a complete set of the Killing vectors $\{\xi ^{(i)}|\,i=1,\ldots,10\} $, for an each $p$-form $F_{(p)}$, we have to solve the set of
equations
\begin{equation*}
\pounds _{\xi ^{(i)}}F_{\mu_1\cdots \mu_p}\equiv \partial _{\mu_1}\xi ^{(i)\nu }F_{\nu \mu _2\cdots \mu_p}+\partial _{\mu_2}\xi
^{(i)\nu }F_{\mu_1\nu \mu_3 \cdots \mu_p}+\cdots +\partial _{\mu_{p}}\xi ^{(i)\nu} F_{\mu _1\cdots \mu_{p-1}\nu }+\xi ^{(i)\nu }
\partial_\nu F_{\mu_1\cdots \mu_{p}}=0\text{\thinspace}.
\end{equation*}
It is straighforward to show that components of a $p$-form invariant under the Lie-dragging along the vectors $\partial _t$, $\partial _\varphi$
and $\partial _\alpha$, do not depend on the coordinates $t$, $\varphi$ and $\alpha$.

In particular, we are interested in a $3$-form
\begin{equation}
F_{(3)}=F_{t\rho \varphi }(\rho )\,dt\wedge d\rho \wedge d\varphi +F_{\psi\theta \alpha }(\psi ,\theta )\,d\psi \wedge d\theta \wedge d\alpha \,.
\end{equation}
Invariance under $\xi ^{(2)}$ gives
\begin{equation}
\pounds _{\xi ^{(2)}}F_{t\rho \varphi }=-\rho \partial _{\rho }F_{t\rho \varphi }=0\quad \Rightarrow \quad F_{t\rho \varphi }=e_1=Const\,,
\end{equation}
and invariance under the action of $\xi ^{(5)}$ leads to
\begin{equation}
\pounds _{\xi ^{(5)}}F_{\psi \theta \alpha }=\cos \alpha \left( \cot \theta
F_{\psi \theta \alpha }-\partial _{\theta }F_{\psi \theta \alpha }\right)
=0\quad\Rightarrow \quad F_{\psi \theta \alpha }(\psi ,\theta )=A(\psi )\sin \theta \,.
\end{equation}
Finally, the $\xi ^{(8)}$-invariance gives
\begin{equation}
\pounds _{\xi ^{(8)}}F_{\psi \theta \alpha }=\sin \theta \cos \theta \left(2A\cot \psi -\partial _{\psi }A\right) =0
\quad \Rightarrow \quad A(\psi)=p_1\sin ^2\psi \,,
\end{equation}%
where $p_1=Const.$ All other Killing vectors leave this solution for $F_{\psi \theta \alpha }$ invariant.

Therefore, the most general $3$-form possessing the same isometries as the metric $G_{\mu \nu }$ is
\begin{equation}
F_{(3)}=e_1\,dt\wedge d\rho \wedge d\varphi +p_1\varepsilon (S^3 )\,,
\label{F3electric}
\end{equation}
where $\varepsilon (S^3 )=\sin ^2\psi \sin \theta \,d\psi \wedge d\theta\wedge d\alpha $.


\begin{thebibliography}{9}

\bibitem{arXiv:0905.2612}
  D.~Anninos, M.~Esole and M.~Guica,
  ``Stability of warped AdS$_3$ vacua of topologically massive gravity,''
  JHEP\ {\bf 0910}, 083  (2009)
  [arXiv:0905.2612 [hep-th]].

\bibitem{arXiv:0807.3040}
  D.~Anninos, W.~Li, M.~Padi, W.~Song and A.~Strominger,
  ``Warped AdS$_3$ Black Holes,''
  JHEP\ {\bf 0903}, 130  (2009)
  [arXiv:0807.3040 [hep-th]].

\bibitem{hep-th/9204099}
  M.~Banados, C.~Teitelboim and J.~Zanelli,
  ``The Black hole in three-dimensional space-time,''
  Phys.\ Rev.\ Lett.\ \ {\bf 69}, 1849  (1992)
  [hep-th/9204099].

\bibitem{gr-qc/9302012}
  M.~Banados, M.~Henneaux, C.~Teitelboim and J.~Zanelli,
  ``Geometry of the $(2+1)$ black hole,''
  Phys.\ Rev.\ D\ {\bf 48}, 1506  (1993)
  [gr-qc/9302012].

\bibitem{Guica:2008mu}
  M.~Guica, T.~Hartman, W.~Song and A.~Strominger,
  ``The Kerr/CFT Correspondence,''
  Phys.\ Rev.\ D {\bf 80}, 124008 (2009)
  [arXiv:0809.4266 [hep-th]].

\bibitem{arXiv:1103.2355}
  I.~Bredberg, C.~Keeler, V.~Lysov and A.~Strominger,
  ``Cargese Lectures on the Kerr/CFT Correspondence,''
  Nucl.\ Phys.\ Proc.\ Suppl.\ \ {\bf 216}, 194  (2011)
  [arXiv:1103.2355 [hep-th]];\\
  G.~Compere,
  ``The Kerr/CFT correspondence and its extensions: a comprehensive review,''
  arXiv:1203.3561 [hep-th].


\bibitem{arXiv:0808.1912}
  G.~Compere, S.~Detournay and M.~Romo,
  ``Supersymmetric Godel and warped black holes in string theory,''
  Phys.\ Rev.\ D\ {\bf 78}, 104030  (2008)
  [arXiv:0808.1912 [hep-th]].

\bibitem{Colgain:2010rg}
  E.~O. Colgain and H.~Samtleben,
  ``3D gauged supergravity from wrapped M5-branes with AdS/CMT applications,''
  JHEP {\bf 1102}, 031 (2011)
  [arXiv:1012.2145 [hep-th]].

\bibitem{Duff:1998cr}
  M.~J.~Duff, H.~Lu and C.~N.~Pope,
  ``AdS$_3\times S^3$ (un)twisted and squashed, and an $O(2,2;\mathbb{Z})$ multiplet of
  dyonic strings,''
  Nucl.\ Phys.\  B {\bf 544}, 145 (1999)
  [arXiv:hep-th/9807173].

\bibitem{Bena:2012wc}
  I.~Bena, M.~Guica and W.~Song,
  ``Un-twisting the NHEK with spectral flows,''
  arXiv:1203.4227 [hep-th].

\bibitem{Anninos:2008qb}
  D.~Anninos,
  ``Hopfing and Puffing Warped Anti-de Sitter Space,''
  JHEP {\bf 0909}, 075 (2009)
  [arXiv:0809.2433 [hep-th]].


\bibitem{noi}
Work in progress.

\bibitem{ef3}
D.~Astefanesei, K.~Goldstein, R.~P.~Jena, A.~Sen and S.~P.~Trivedi,
  ``Rotating attractors,''
  JHEP\ {\bf 0610}, 058  (2006)
  [hep-th/0606244].

\bibitem{Cremmer:1997ct}
  E.~Cremmer, B.~Julia, H.~Lu and C.~N.~Pope,
  ``Dualization of dualities. 1.,''
  Nucl.\ Phys.\ B {\bf 523}, 73 (1998)
  [hep-th/9710119].

\bibitem{ef1}
A.~Sen,
  ``Black hole entropy function and the attractor mechanism in higher derivative gravity,''
  JHEP\ {\bf 0509}, 038  (2005)
  [hep-th/0506177].

\bibitem{ef2}
A.~Sen,
  ``Entropy function for heterotic black holes,''
  JHEP\ {\bf 0603}, 008  (2006)
  [hep-th/0508042].

\bibitem{Dabholkar:2006tb}
    A.~Dabholkar, A.~Sen and S.~P.~Trivedi,
    ``Black hole microstates and attractor without supersymmetry,''
    JHEP {\bf 0701}, 096 (2007)
    [arXiv:hep-th/0611143].

\bibitem{arXiv:1008.3852}
  D.~Astefanesei, N.~Banerjee and S.~Dutta,
  ``Moduli and electromagnetic black brane holography,''
  JHEP\ {\bf 1102}, 021  (2011)
  [arXiv:1008.3852 [hep-th]].


\bibitem{arXiv:1104.4121}
D.~Astefanesei, N.~Banerjee and S.~Dutta,
  ``Near horizon data and physical charges of extremal AdS black holes,''
  Nucl.\ Phys.\ B\ {\bf 853}, 63  (2011)
  [arXiv:1104.4121 [hep-th]].


\bibitem{arXiv:0708.1270}
  A.~Sen,
  ``Black Hole Entropy Function, Attractors and Precision Counting of Microstates,''
  Gen.\ Rel.\ Grav.\ \ {\bf 40}, 2249  (2008)
  [arXiv:0708.1270 [hep-th]].

\bibitem{Astefanesei:2005yj}
  D.~Astefanesei, R.~B.~Mann and C.~Stelea,
  ``Nuttier bubbles,''
  JHEP {\bf 0601}, 043 (2006)
  [hep-th/0508162].

\bibitem{Astefanesei:2006sy}
  D.~Astefanesei, K.~Goldstein and S.~Mahapatra,
  ``Moduli and (un)attractor black hole thermodynamics,''
  Gen.\ Rel.\ Grav.\  {\bf 40}, 2069 (2008)
  [arXiv:hep-th/0611140].

\bibitem{arXiv:1009.5039}
  M.~Guica and A.~Strominger,
  ``Microscopic Realization of the Kerr/CFT Correspondence,''
  JHEP\ {\bf 1102}, 010  (2011)
  [arXiv:1009.5039 [hep-th]] ;\\
S.~El-Showk and M.~Guica,
  ``Kerr/CFT, dipole theories and nonrelativistic CFTs,''
  arXiv:1108.6091 [hep-th];\\
W.~Song and A.~Strominger,
  ``Warped AdS3/Dipole-CFT Duality,''
  arXiv:1109.0544 [hep-th];\\
G.~Compere, W.~Song and A.~Virmani,
  ``Microscopics of Extremal Kerr from Spinning M5 Branes,''
  JHEP {\bf 1110}, 087 (2011)
  [arXiv:1010.0685 [hep-th]].

\bibitem{Astefanesei:2009sh}
  D.~Astefanesei and Y.~K.~Srivastava,
  ``CFT Duals for Attractor Horizons,''
  Nucl.\ Phys.\ B {\bf 822}, 283 (2009)
  [arXiv:0902.4033 [hep-th]].

 \bibitem{hep-th/0504231}
  S.~Detournay, D.~ Orlando, M.~Petropoulos and Ph.~Spindel,
  ``Three-dimensional black holes from deformed anti-de Sitter,''
  JHEP {\bf 0507}, 072 (2005)
  [hep-th/0504231].


\end{thebibliography}
\end{document}